\documentclass[a4paper,10pt,twoside]{cpc-hepnp}
\usepackage{CJK,upgreek,fancyhdr}
\usepackage{multicol}
\usepackage{graphicx}
\usepackage{booktabs}
\usepackage{amssymb,bm,mathrsfs,bbm,amscd}
\usepackage[tbtags]{amsmath}
\usepackage{lastpage}
\usepackage{longtable}
\usepackage{changes}
\usepackage{lineno,hyperref}
\usepackage{threeparttable} 
\usepackage{multirow}
\usepackage{subfigure} 
\usepackage{ulem}
\usepackage{color}
\usepackage{diagbox}
\usepackage{verbatim}
\usepackage[font=small,labelfont=bf]{caption}
\newenvironment{figurehere}{\def\@captype{figure}}{}
\newenvironment{tablehere}{\def\@captype{table}}{}
\makeatother
\begin{document}

\begin{CJK*}{GB}{gbsn}

\fancyhead[c]{\small Chinese Physics C~~~Vol. xx, No. x (202x) xxxxxx}
\fancyfoot[C]{\small 010201-\thepage}

\title{Theoretical calculations of proton emission half-lives based on a deformed Gamow-like model
\thanks{This work is supported in part by the National Natural Science Foundation of China (Grants No. 12175100 and No. 11975132), the Construct Program of the Key Discipline in Hunan Province, the Research Foundation of Education Bureau of Hunan Province, China (Grant No. 21B0402 and No. 18A237), the Natural Science Foundation of Hunan Province, China (Grants No. 2018JJ2321), the Innovation Group of Nuclear and Particle Physics in USC, the Shandong Province Natural Science Foundation, China (Grant No. ZR2022JQ04), the Hunan Provincial Innovation Foundation For Postgraduate (Grant No. CX20220993) and the Opening Project of Cooperative Innovation Center for Nuclear Fuel Cycle Technology and Equipment, University of South China (Grant No. 2019KFZ10).}}
\author{%
\quad Dong-Meng Zhang (张冬萌)$^{1}$	
\quad Xiao-Yuan Hu (胡笑源)$^{1}$
\quad Lin-Jing Qi 
(亓林静)$^{1}$
\quad Hong-Ming Liu 
(刘宏铭)$^{2}$\\
\quad Ming Li
(李明)$^{1;1)}$\email{liming1631223@163.com}
\quad Xiao-Hua Li
(李小华)$^{1,3,4,5;2)}$\email{lixiaohuaphysics@126.com}\\
}
\maketitle
\address{%
 \begin{flushleft}
$^1$School of Nuclear Science and Technology, University of South China, Hengyang 421001, China\\
$^2$Institute of Modern Physics, Fudan University, Shanghai 200433, China\\
$^3$National Exemplary Base for International Sci \& Tech. Collaboration of Nuclear Energy and Nuclear Safety,
University of South China, Hengyang 421001, China\\
$^4$Cooperative Innovation Center for Nuclear Fuel Cycle Technology \& Equipment, University of South China, Hengyang 421001, China\\
$^5$Key Laboratory of Low Dimensional Quantum Structures and Quantum Control, Hunan Normal University, Changsha 410081, China\\
 \end{flushleft}
}
\begin{abstract}
In the present study, proton emission half-lives have been investigated for the deformed proton emitters with $53\leq Z \leq 83$ in the deformed Gamow-like model, where the deformation effect has been included in the Coulomb potential. The experimental half-lives of proton emitters can be reproduced within a factor of 3.45. For comparison, other results from the universal decay law and the new Geiger-Nuttall law are presented as well. Furthermore, the relevance of the half-lives to the angular momentum $l$ for $^{117}$La, $^{121}$Pr, $^{135}$Tb and $^{141}$Ho has been analyzed, and corresponding possible values of $l$ has been put forward: $l=$3, 3, 4, 4. 
\end{abstract}


\footnotetext[0]{\hspace*{-3mm}\raisebox{0.3ex}{$\scriptstyle\copyright$}2023
Chinese Physical Society and the Institute of High Energy Physics
of the Chinese Academy of Sciences and the Institute
of Modern Physics of the Chinese Academy of Sciences and IOP Publishing Ltd}%
	
\begin{multicols}{2}
	
\section{Introduction}
\label{Sec.I}
Proton emission is one of the foremost decay modes of proton-rich nuclei distant from the $\beta$-stability line \cite{C. Xu Phys. Rev. C 2016}. The proton-rich nuclei have positive $Q_p$ values for proton emissions with a spontaneous trend to release excess protons \cite{Basu D N Phys. Rev. C 2005,M. Balasubramaniam Phys. Rev. C 2005}. Proton emission is a spontaneous nuclear reaction that can be expressed as
\begin{eqnarray}
^A_Z{(X)}\rightarrow^{A-1}_{Z-1}{(Y)}+^1_1(H)+Q_p.
\label{eq1}
\end{eqnarray} 
In the isomeric state of $^{53}{\rm Co}$, Jackson \emph{et al.} \cite{Jackson K Phys. Lett. B 1970,Cerny J Phys. Lett. B 1970} made the first observation of proton emission in 1970. So far, proton emissions have been discovered from 45 nuclei with $53\leq Z \leq 83$ in the ground state or isomeric state \cite{Qian Y and Ren Eur. Phys. J. A 2016,Kondev F G Chin. Phys. C 2021,Delion D S Phy. Rev. Lett. 2006,Blank B Prog. Part. Nucl. Phys. 2008}. Studying on the proton emission may have important implications for facilitating the extant nuclear theories and models, obtaining spectroscopic information, and determining nuclear deformation, etc \cite{Zhang H F J. Phys. G: Nucl. Part. Phys. 2010,Chen J L J. Phys. G: Nucl. Part. Phys. 2019,Delion D S Phys. Rev. C 2009,Karny M Phys. Lett. B 2008,Zhang Z X Chin. Phys. C 2018,Zhu D X et al Phys. Scr. 2022,Xing F Z et al Chin. Phys. C 2021}.  Therefore, there have been numerous attempts to study this phenomenon in theory \cite{Hofmann S Z. Phys. A 1982,O. Klepper Z. Phys. A: At. Nucl. 1982,T. Faestermann Phys. Lett. B 1984,S. Hofmann 1984,C. N. Davids Phys. Rev. Lett. 1998,Santhosh K P Pramana J. Phys.,Z. Y. Yuan Sci. China Phys. Mech. Astron. 2023,Y. Y. Xu Nucl. Sci. Tech. 2023,D. M. Zhang Nucl. Sci. Tech. 2023}.

Since most proton emitters are spherical or moderately deformed, the calculations of proton emission half-lives are usually simplified by assuming a spherical shape for the daughter nucleus. Furthermore, the Wentzel-Kramers-Brillouin (WKB) approximation is capable of handling proton emission because it shares the same physical processes as $\alpha$ decay \cite{S. Luo Commun. Theor. Phys. 2023,D. D. Ni Phys. Rev. C 2008,T. K. Dong Phys. Rev. C 2008,Y. Y. Xu Chin. Phys. C 2022}, cluster radioactivity \cite{L. J. Qi Phys. Rev. C 2023,L. J. Qi Chin. Phys. C 2023,D. N. Poenaru Phys. Rev. C 2011,Y. Z. Wang Chin. Phys. C 2021}, and two-proton emission \cite{L. Zhou Nucl. Sci. Tech. 2022,D. Q. Fang Chin. Sci. Bull. 2020,J. P. Cui Phys. Rev. C 2020,H. M. Liu Chin. Phys. C 2021}, which can be handled through barrier penetration. Based on the WKB approximation, various calculations for spherical proton emitters with different models or potentials can obtain similar calculated values, which closely match the experimental half-lives \cite{Zhang H F Sci. China. Ser. G:Phy. Mech. Astron.,Bhattacharya M Phys. Lett. B 2007,Dong J M Chin. Phys. C 2010,Zdeb A Eur. Phys. J. A 2016,Guo C Eur. Phys. J. A 2014,Guo C Nucl. Phys. A 2013,Dong J M Phys. Rev. C 2009,Wang Y Z Phys. Rev. C 2017,J. L. Chen Eur. Phys. J. A 2021,Y. Q. Xin 2021,Y. B. Qian Chin. Phys. Lett. 2010-2}. Although there is good agreement between experimental data and the calculated ones, these spherical models can be further improved after considering the effect of nuclear deformation, leading to a more microscopic understanding of proton emission. Nowadays, various theoretical methods are used to investigate the proton emission of deformed proton emitters, such as Coulomb and proximity potential model for deformed nuclei (CPPMDN) \cite{Santhosh K P Phys. Rev. C 2017}, deformed density-dependent model (D-DDM) with the single-folding potential \cite{Qian Y and Ren Eur. Phys. J. A 2016,Y. B. Qian Chin. Phys. Lett. 2010}, modified unified fission model (M-UFM) with the deformation-dependent Coulomb potential \cite{N. S. Rajeswari Eur. Phys. J. A 2014}, modified two-potential approach for deformed nuclei (D-TPA) \cite{D. M. Zhang Phys. Rev. C 2023,J. H Cheng Phys. Rev. C 2022}, deformation dependent screened decay law (D-SDL) \cite{R. Budaca Nucl. Phys. A 2022}, and others. These theoretical calculations enhance our knowledge of the proton emission phenomena and offer reasonable estimates for the proton emission half-life. 

In 2013, based on the Gamow theory, Zdeb \emph{et al.} proposed a phenomenological model for investigating the half-lives of $\alpha$ decay and cluster radioactivity named as Gamow-like model (GLM) \cite{A. Zdeb Phys. Rev. C 2013}. In this model, the outside potential always corresponds to the Coulomb potential, while the interior potential is depicted as a square well. Subsequently, considering the centrifugal barrier's effects, GLM was successfully extended to describe the proton emission \cite{A. Zdeb Eur. Phys. J. A 2016}.  
Furthermore, Chen \emph{et al.} \cite{Chen J L J. Phys. G: Nucl. Part. Phys. 2019} modified the Gamow-like model (M-GLM) by introducing a screened electrostatic barrier named as Hulth\'en potential as well as calculated the proton emission half-lives. Recently, GLM has been widely applied in evaluating the two-proton emission half-lives \cite{H.M. Liu Chin. Phys. C 2021,H.M. Liu  Int. J. Mod. Phys. E 2021,D. X. Zhu Nucl. Sci. Tech. 2022}. The results obtained by GLM and M-GLM can reproduce the experimental half-lives for the spherical nuclei to a substantial degree. However, the nuclear deformation effect is consequently crucial in calcutions of proton emission \cite{Qian Y and Ren Eur. Phys. J. A 2016,Y. B. Qian Chin. Phys. Lett. 2010}. Extending the model from the spherical example to a deformed form would be intriguing. In this work, taking into consideration the deformation effect of Coulomb-type potential, we modified the GLM proposed by Zdeb \emph{et al.} \cite{A. Zdeb Eur. Phys. J. A 2016}, denoted as the D-GLM, to calculate the proton emission half-lives of deformed nuclei with $53\leq Z \leq 83$. It turns out that our calculations and experimental values correspond in a good manner. 

The structure of this article is as follows. The deformed Gamow-like model's theoretical foundation is outlined in depth in Section \ref{Sec.II}. Section \ref{Sec.III} presents the results and related discussion. Finally, Section \ref{Sec.IV} provides a succinct summary. 

\section{Theoretical framework}
\label{Sec.II}
In the D-GLM, the proton emission half-life $T_{1/2}$ is correlated with the decay constant $\lambda_p$ and can be written as
\begin{eqnarray}
T_{1/2}=\frac{\ln2}{\lambda_p}.
\label{eq2}
\end{eqnarray}
$\lambda_p$ is defined as
\begin{eqnarray}
{\lambda_p}=S_p\nu P,
\label{eq3}
\end{eqnarray}
where $S_p$ is the spectroscopic factor of the emitted proton-daughter system. $\nu$ represents the assault frequency related to the harmonic oscillation frequency presented in Nilsson potential. It might be described as \cite{S.G.Nilsson}
\begin{eqnarray}
h\nu=\hbar\omega\simeq \frac{41}{A^{1/3}},
\label{eq4}
\end{eqnarray}
where $h$, $\hbar$, $\omega$, and $A$ denote the Planck constant, reduced Planck constant, angular frequency, and mass number of the proton emitter, respectively.

On the basis of the semi-classical WKB approximation, the barrier penetrability $P$ of the proton penetrating the external barrier can be determined.  By averaging $P_{\theta}$ in all directions while taking the impact of deformation into account, we can determine the overall penetration probability. Consequently, $P$ is expressed as
\begin{eqnarray}
P=\frac{1}{2}\int_0^{\pi}P_{\theta}\sin\theta{d\theta},
\label{eq5}
\end{eqnarray}
where $\theta$ represents the angle formed by the daughter nucleus's symmetry axis and the radius vector. $P_{\theta}$ is the polar-angle-dependent penetration probability of proton emission. It can be given by
\begin{eqnarray}
P_{\theta}={\rm{exp}}\left[-\frac{2}{\hbar}\int_{R_{\rm in}(\theta)}^{R_{\rm out}(\theta)}\sqrt{2\mu \vert V(r,\theta)-Q_p \vert }{dr}\right],
\label{eq6}
\end{eqnarray}
where $\mu=m_p m_d/(m_p+m_d)\approx938.3{\tiny }\times A_d/A\ {\rm MeV/c^2}$ denotes the reduced mass. $m_p$, $m_d$, $A_d$ and $A$ are the mass of the emitted proton, daughter nucleus, the mass number of daughter nucleus, and proton emitter, respectively. For the released energy $Q_p$, it contains the electrostatic screening effect and can be given by
\begin{eqnarray}
Q_p=\Delta M-(\Delta M_d+\Delta M_{p})+k(Z^{\varepsilon}-Z_d^{\varepsilon}),
\label{eq7}
\end{eqnarray}
where $\Delta M, \Delta M_d$ and $\Delta M_p$ represent the mass excesses of parent nucleus, daughter nucleus and the emitted proton, respectively. The relevant data are derived from the most current atomic mass table \cite{Kondev F G Chin. Phys. C 2021}. The last term $k(Z^{\varepsilon}-Z_d^{\varepsilon})$ denotes the screening effect of atomic electrons, corresponding to $k=13.6$ eV, $\varepsilon=2.408$ for $Z<60$, and $k=8.7$ eV, $\varepsilon=2.517$ for $Z\geq 60$ \cite{Denisov V Y Phys. Rev. C 2005,Huang K N At. Data Nucl. Data Tables 1976}. $Z$ and $Z_d$ are the proton numbers of parent nuclei and daughter nuclei, respectively.

In addition, $V(r, \theta)$ is the total interacting potential between the residual daughter nucleus and emitted proton (see Fig. \ref{fig1}), which can be expressed as
\begin{eqnarray}
V(r,\theta) =
\left\{
\begin{array}{ccc}
-V_0,& \quad 0\leq r\leq R_{in},
\\V_{C}(r,\theta) + V_{l}(r),& \quad r>R_{in}.
\end{array}
\right.
\label{eq8} 
\end{eqnarray}
Here, $V_0$ denotes the depth of the inner nuclear potential well with $V_0=25A_p$ MeV, which can be obtained from Ref. \cite{R. Blendowske Phys. Rev. Lett. 1988}. In Eq. \ref{eq6}, $R_{in}(\theta)$ and $R_{out}(\theta)$ are two classical turning points. The outer turning point $R_{out}(\theta)$ is derived by the equation $V(R_{out}(\theta))=Q_p$. $R_{in}(\theta)$ is the inner turning point, which can be obtained by adding the radius of the daughter nucleus and the width of the radial distribution of the emitted proton. It can be given by

\begin{eqnarray}
R_{in}(\theta)=R_p+R_d(\theta),
\label{eq9}
\end{eqnarray}
where $R_p=r_0A_p^{1/3}$ is the radius of the emitted proton with $A_p=1$. $r_0$ is the effective nuclear radius constant, which is the only adjustable parameter in our model. Here, the radius of the daughter nucleus $R_d(\theta)$ is written as 
\begin{eqnarray}
R_d(\theta)=r_0A_d^{1/3}\left[1+\sum_{\lambda}\beta_{\lambda}Y_{\lambda 0}(\theta)\right], 
\label{eq10}
\end{eqnarray}
where $\beta_{\lambda}$ refers to the collection of deformation parameters of the daughter nucleus ($\lambda=2$, 4, 6, correspond to the quadrupole, hexadecapole, and hexacontatetrapole deformations) \cite{A. J. Sierk 2016}. $Y_{\lambda 0}(\theta)$ is a spherical harmonics function.

Furthermore, the Coulomb interaction between the daughter nucleus and emitted proton, with higher multipole deformations included by following $(\lambda=2,4,6)$ \cite{R. K. Gupta J. Phys. G: Nucl. Part. Phys. 2005,C. Y. Wong 1973,A. J. Sierk 2016}, is given as
\begin{equation}
\begin{split}
V_{C}(r,\theta)&=\frac{Z_dZ_pe^2}{r}+3Z_dZ_pe^2\sum_{\lambda}\frac{1}{2\lambda+1}\\&\times \frac{R_d^{\lambda}(\theta)}{r^{\lambda+1}} Y_{\lambda 0}(\theta)\left[\beta_{\lambda}+\frac{4}{7}\beta_{\lambda}^2Y_{\lambda 0}(\theta)\right],
\end{split}
\label{eq11} 
\end{equation}
where $r$ denotes the distance between the daughter nucleus and emitter proton centers.

Here, the centrifugal potential term $V_l(r)$ has been introduced as
\begin{eqnarray}
V_{l}(r)=\frac{\hbar^2l(l+1)}{2\mu r^2},
\label{eq12} 
\end{eqnarray}
where $l$ is the angular momentum carried by the emitted proton, which is determined by the spin-parity conservation rule. It can be expressed as \cite{V. Y. Denisov 2009}
\begin{eqnarray}
l=
\left\{
\begin{array}{ll}
\Delta_j & \quad \mbox{for even}\ \Delta_j \ \mbox{and}\ \pi=\pi_d,\vspace{0.3em}\\
\Delta_j+1 & \quad \mbox{for even}\ \Delta_j \ \mbox{and}\ \pi\neq\pi_d,\vspace{0.3em} \\  
\Delta_j & \quad \mbox{for odd}\ \Delta_j \ \mbox{and}\ \pi\neq\pi_d,\vspace{0.3em} \\
\Delta_j+1 & \quad \mbox{for odd}\ \Delta_j \ \mbox{and}\ \pi=\pi_d,
\end{array}
\right.
\label{eq13} 
\end{eqnarray}
where $\Delta_j=\vert j-j_d-j_p \vert$. $j$, $\pi$, $j_d$, $\pi_d$, and $j_p$, $\pi_p$ denote the spin and parity values of the parent nucleus, daughter nucleus, and emitted proton, respectively.

\begin{figurehere}\centering
	\includegraphics[width=8.5cm]{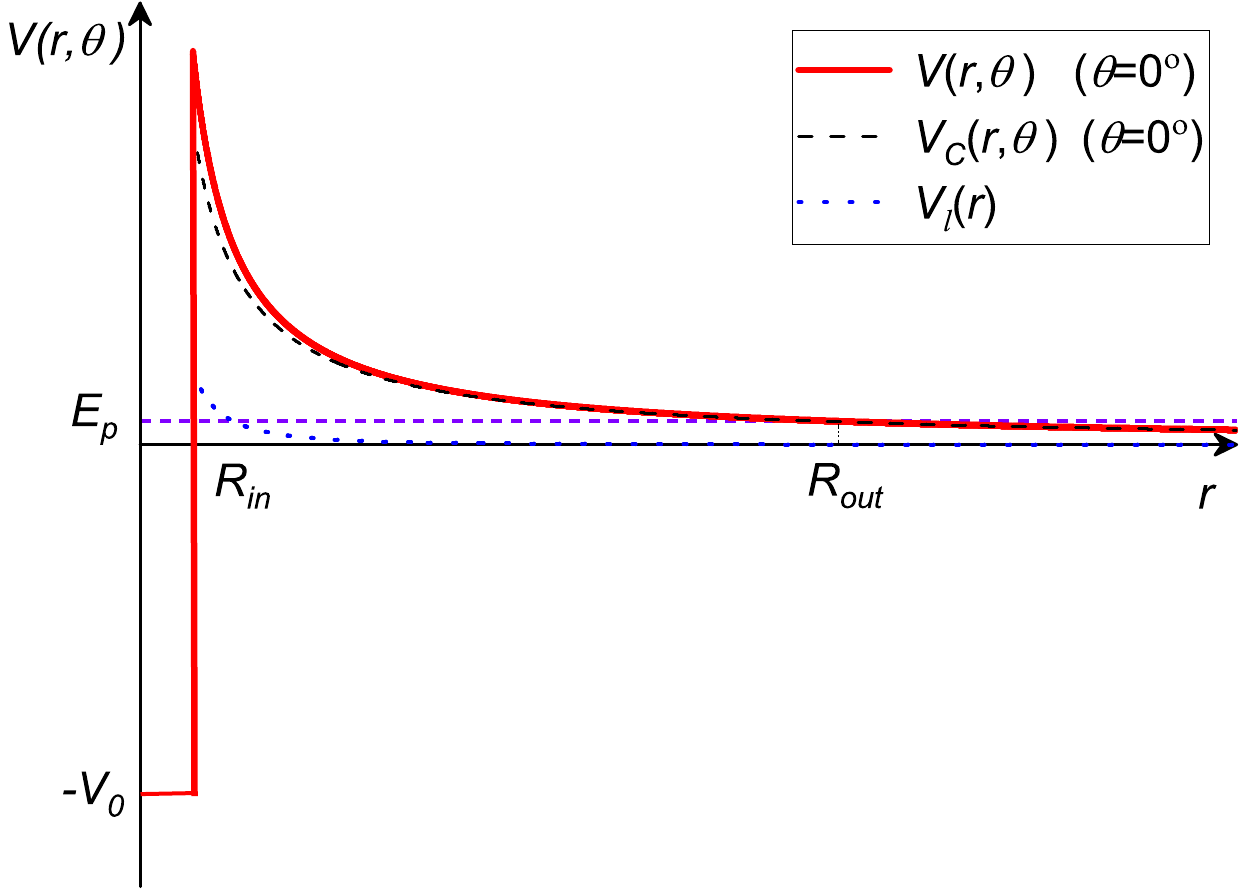}
	\figcaption{(color online) A diagram showing the relationship between the potential energy and the decay system's center-of-mass distance for $^{109}$I. The black and blue dotted lines correspond to the Coulomb potential $V_C(r,\theta)$ ($\theta=0^{\circ}$) and centrifugal potential $V_l(r)$.}
	\label{fig1}
\end{figurehere}

\section{Results and discussion}
\label{Sec.III}
The purpose of this work is to investigate the proton emission half-lives for deformed proton emitters with the proton number range $53\leq Z \leq 83$ from the ground state and/or isomeric state based on the D-GLM. Taking into account the deformation effect of the Coulomb potential, there is an adjustable parameter, i.e., the effective nuclear radius constant ($r_0$) in the D-GLM, which can be obtained by fitting 45 experimental data of the proton emission half-lives. While, the standard deviation $\sigma$ can indicate the deviation between the experimental half-lives and the calculated ones, which is defined as 
\begin{eqnarray}
\sigma=\sqrt{\frac{1}{N}\sum_{i=1}^N{({\rm log}_{10}{T_{1/2}^{{\rm cal}.i}}-{\rm log}_{10}{T_{1/2}^{{\rm exp}.i}})^2}},
\label{eq14}
\end{eqnarray}
where ${\rm log}_{10}{T_{1/2}^{{\rm exp}.i}}$ and ${\rm log}_{10}{T_{1/2}^{{\rm cal}.i}}$ represent the logarithmic form of experimental proton emission half-life and the calculated one for the $i$-th nucleus, respectively. The smallest standard deviation $\sigma=0.533$ can be obtained when the effective nuclear radius constant is taken as $r_0=1.20$ fm (see Fig. \ref{fig2}). This value is in the range of parameter described in Ref. \cite{W. D. Myers Ann. Phys 1991,W. D. Myers Ann. Phys 1974}.

In the following, within the framework of the D-GLM, we systematically investigate the half-lives of 45 deformed proton emitters. For comparison, the calculated results obtained by the Gamow-like model proposed by Xiao \emph{et al.} \cite{Q. Xiao 2023} (GLM-Xiao), universal decay law for proton emission (UDLP) \cite{Qi C Phys. Rev. C 2012}, and the new Geiger-Nuttall law (NG-N) \cite{Chen J L Eur. Phys. J. A 2019} are also presented. 

The UDLP was proposed by Qi \emph{et al.} \cite{Qi C Phys. Rev. C 2012} for investigating proton radioactivity, which can be expressed as
\begin{eqnarray}
{\rm log}_{10}{T}_{1/2}=a\chi^{\prime}+b\rho^{\prime}+c+d(l+1)l/\rho^{\prime},
\label{eq15}
\end{eqnarray}
where the adjustable parameters are $a=0.386, b=-0.502, c=-17.8$ and $d=2.386$, respectively. The middle parameters are $\chi^{\prime}=Z_pZ_d\sqrt{\frac{A_pA_d}{(A_p+A_d)Q_p}}$ and $\rho^{\prime}=\sqrt{\frac{Z_pZ_dA_pA_d(A_p^{1/3}+A_d^{1/3})}{A_p+A_d}}$.

The NG-N was proposed by Chen \emph{et al.} \cite{Chen J L Eur. Phys. J. A 2019}, which was a two-parameter empirical formula. It can be expressed as
\begin{eqnarray}
{\rm log}_{10}{T}_{1/2}=a(Z_d^{0.8}+l)Q_p^{-1/2}+b,
\label{eq16}
\end{eqnarray}
where the two parameters $a=0.843$ and $b=-27.194$ were obtained by fitting 44 experimental data of the proton radioactivity half-lives.

All specific calculated half-lives in logarithmic form have been presented in Table. \ref{tab1}. In this table, the first five columns denote proton emitters, proton emission released energy $Q_p$, the spin and parity transition, the angular momentum $l$ taken away by the emitted proton, and the spectroscopic factor $S_p$, respectively. The values of $S_p$ are derived from the relativistic mean field theory (RMF) with the Bardeen-Cooper-Schriffer (BCS) theory \cite{Qian Y and Ren Eur. Phys. J. A 2016} expect for $^{108}$I, $^{144}$Tm, $^{159}$Re$^m$, $^{170}$Au, $^{170}$Au$^m$, $^{176}$Tl, which are taken from Ref. \cite{D. M. Zhang Phys. Rev. C 2023}. The correlative nuclear deformation parameters are taken from Ref. \cite{A. J. Sierk 2016}. The last five columns denote the experimental proton emission half-lives, theoretical ones calculated using the D-GLM, GLM-Xiao, UDLP, and NG-N, respectively. It is worth noting that the present results with our model are basically consistent with the experimental data. 

\begin{figurehere}\centering
	\includegraphics[width=8.5cm]{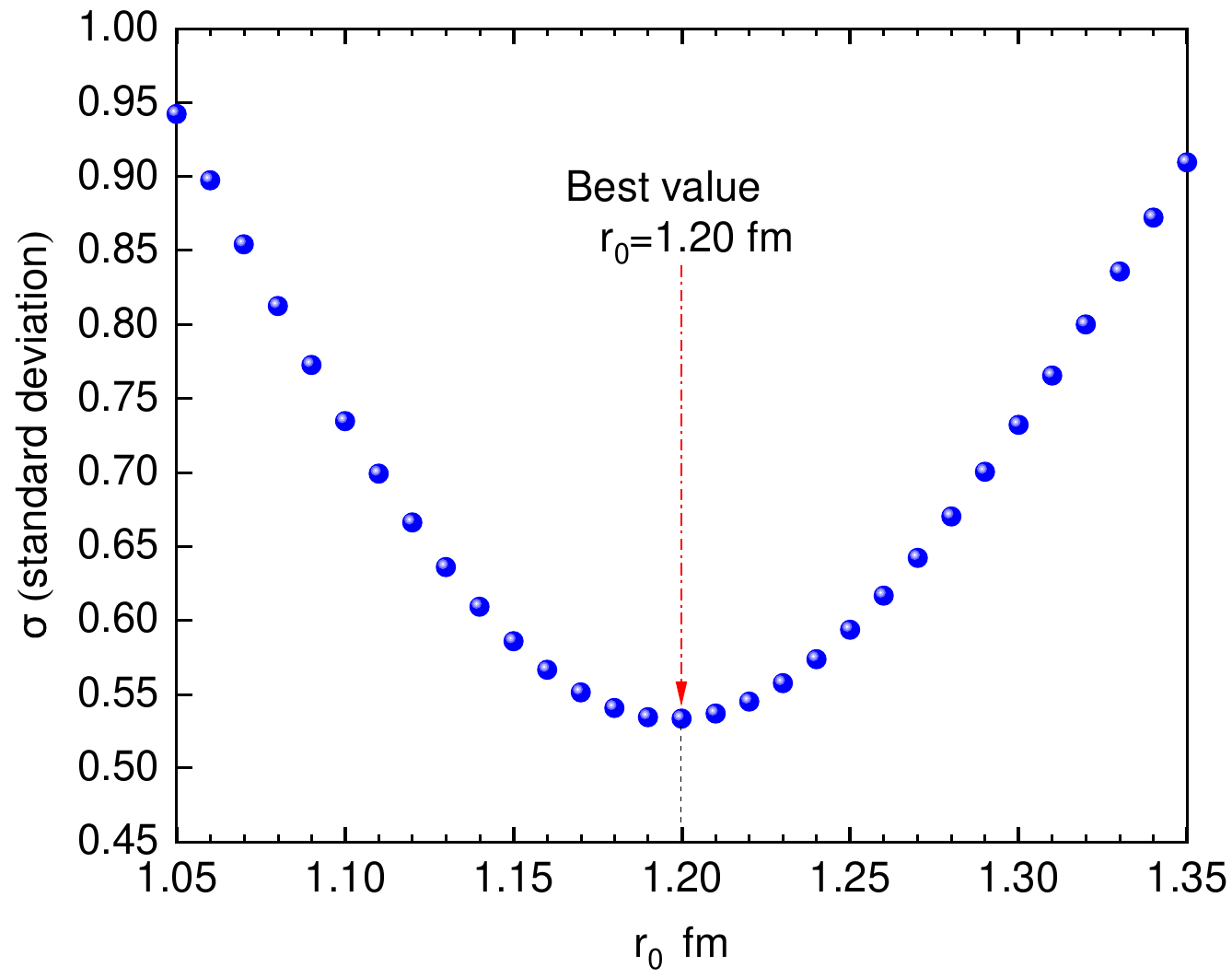}
	\figcaption{(color online) The relevance between the standard deviation $\sigma$ and the value of the parameter $r_0$.}
	\label{fig2}
\end{figurehere}
\begin{figurehere}\centering
	\includegraphics[width=8.5cm]{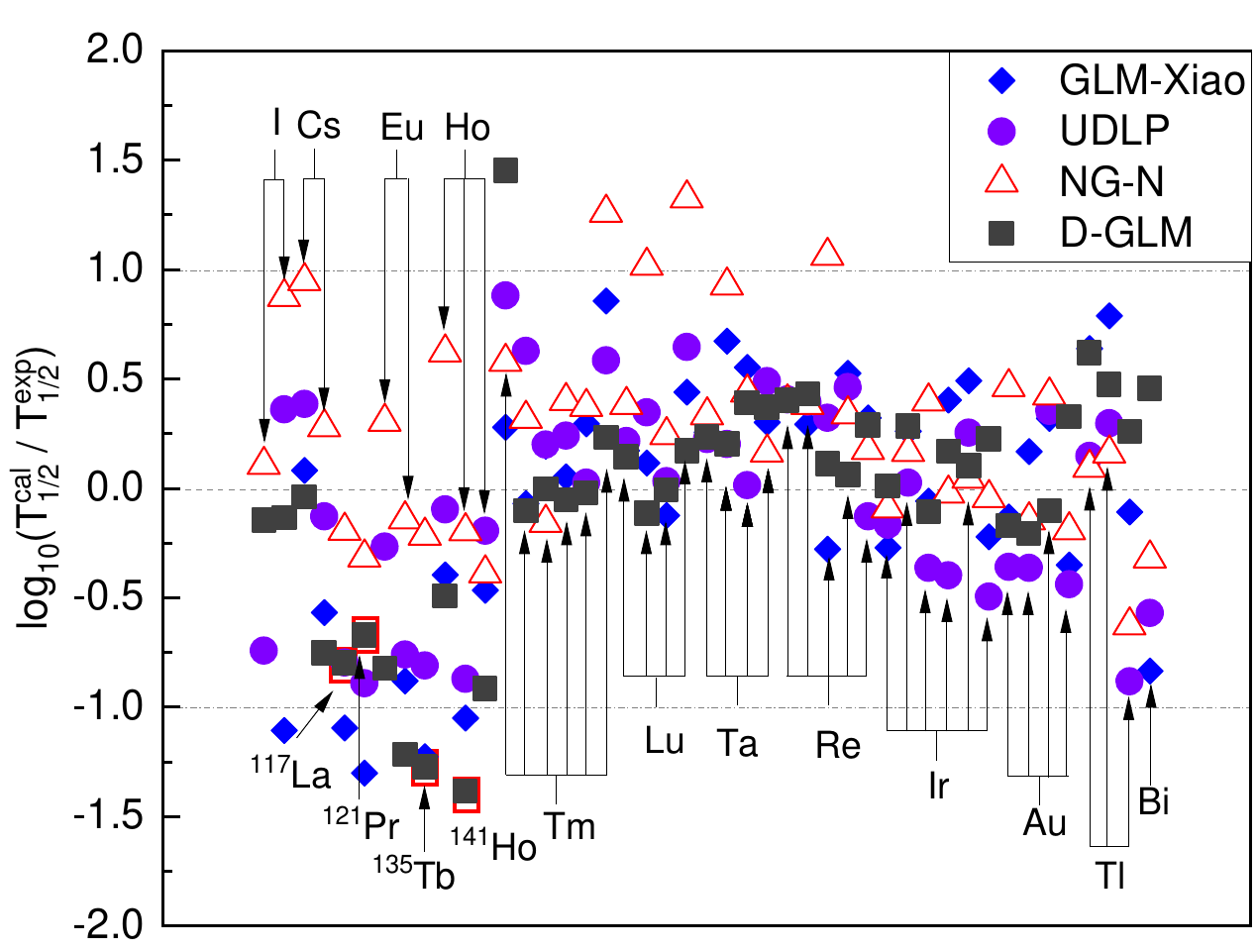}
	\figcaption{(color online) The deviations between the experimental proton emission half-lives and calculated ones in logarithmic form for deformed nuclei. The blue rhombuses, purple circles, red triangles, and black squares correspond to the deviations determined by applying the GLM-Xiao, UDLP, NG-N, and D-GLM, respectively.}
	\label{fig3}
\end{figurehere}
\begin{figure*}[htb]\centering
	\includegraphics[width=\linewidth]{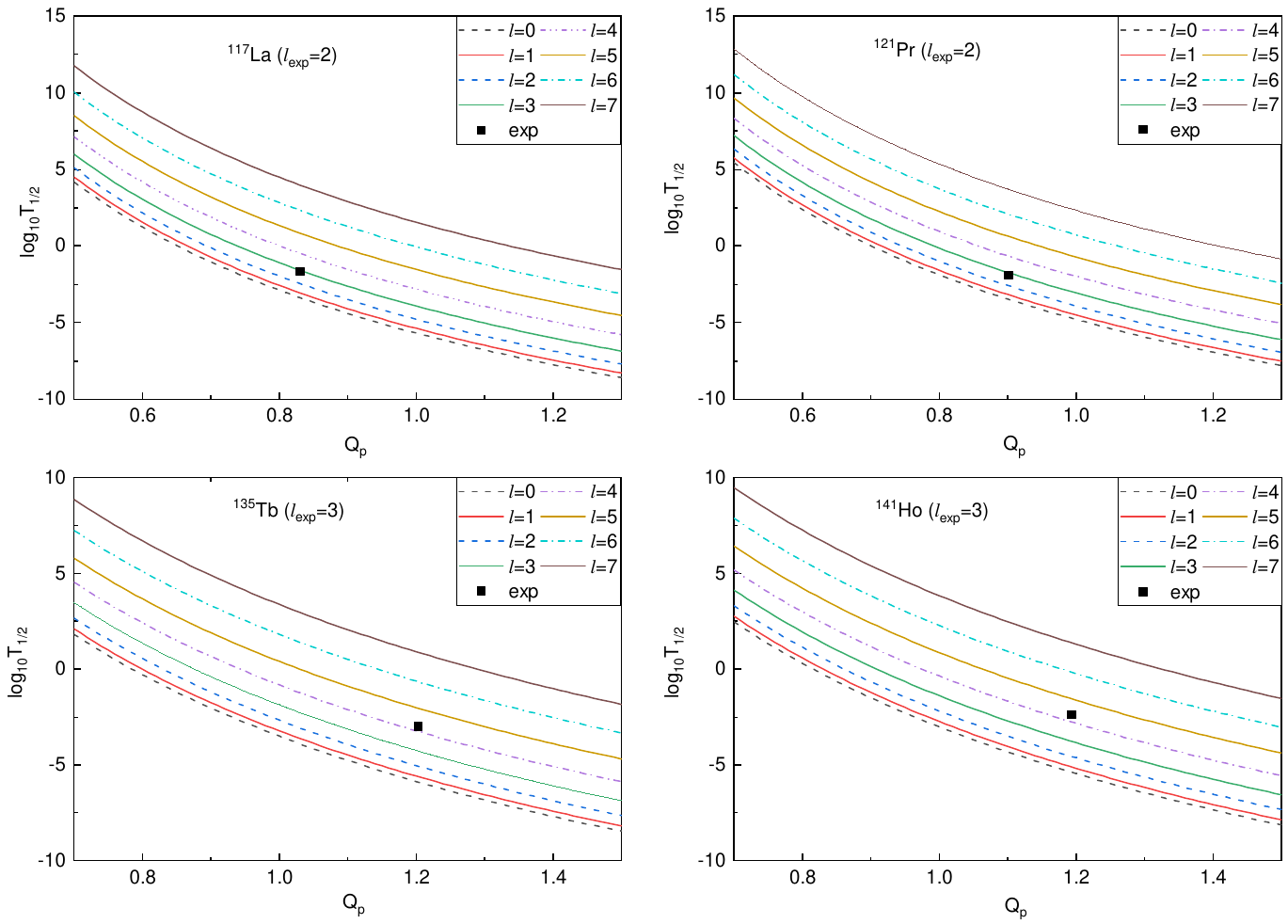}
	\figcaption{(color online) The values ${\rm log}_{10}T_{1/2}$ as functions of $Q_p$ for $^{117}$La, $^{121}$Pr, $^{135}$Tb, and $^{141}$Ho when $l$ is taken as different value.}
	\label{fig4}
\end{figure*}

\begin{table*}[htb]
	\caption{Comparison of experimental proton emission half-lives with the calculated ones by using different theoretical models and/or formulae. The symbol $m$ represents the first isomeric state. The experimental proton emission half-lives, spin and parity are taken from Ref. \cite{Kondev F G Chin. Phys. C 2021}. The released energy are given by Eq. \ref{eq7} with the exception of $Q_p$ value for $^{130}$Eu, $^{159}$Re, $^{161}$Re$^{m}$, $^{164}$Ir and $^{177}$Tl$^{m}$ taken from Refs.\cite{Blank B Prog. Part. Nucl. Phys. 2008,G. Audi Chin. Phys. C 2017}. The values of $S_p$ are derived from Ref. \cite{Qian Y and Ren Eur. Phys. J. A 2016} expect for $^{108}$I, $^{144}$Tm, $^{159}$Re$^m$, $^{170}$Au, $^{170}$Au$^m$, $^{176}$Tl, which are taken from Ref. \cite{D. M. Zhang Phys. Rev. C 2023}. The correlative nuclear deformation parameters are taken from Ref. \cite{A. J. Sierk 2016}.}
	\renewcommand\arraystretch{1}
	\label{tab1}
	\begin{tabular*}{19cm} {@{\extracolsep{\fill}} lccccccccc}
		\hline
		{Nucleus}&$Q_p$&$j^{\pi}\rightarrow j_d^{\pi}$&$l$&$S_p$&&&$\rm{log}_{10}{\emph{T}}_{1/2}$(s)&&
		\\
		\cline{6-10}
		$^{A}Z$&MeV&&&&Exp&D-GLM&GLM-Xiao\cite{Q. Xiao 2023}&UDLP \cite{Qi C Phys. Rev. C 2012}&NG-N \cite{Chen J L Eur. Phys. J. A 2019}\\
		\hline	
		$	^	{	108	}	$I	&	0.610 	&$(1^+)\#\rightarrow5/2^+\#$&	2&	0.089&	0.723 	&	0.578	&&	--0.019 	&	0.829 	\\
		$	^	{	109	}	$I	&	0.829 	&$(3/2^+)\rightarrow0^+$&	2	&0.435&	--4.032 	&	--4.168	&--5.138&	--3.671 	&	--3.157 	\\
		$	^	{	112	}	$Cs	&	0.820 	&$1^+\#\rightarrow5/2^+\#$&	2	&0.464&	--3.310 	&	--3.349	&--3.228&	--2.923 	&	--2.362 	\\
		$	^	{	113	}	$Cs	&	0.981 	&$(3/2^+)\rightarrow0^+$&	2	&0.446&	--4.771 	&	--5.525	&--5.338&	--4.899 	&	--4.490 	\\
		$	^	{	117	}	$La	&	0.831 	&$(3/2^+)\rightarrow0^+$&	2	&0.183&	--1.664 	&	--2.461	&--2.757&	--2.459 	&	--1.857 	\\
		$	^	{	121	}	$Pr	&	0.901 	&$(3/2^+)\rightarrow0^+$&	2	&0.107&	--1.921 	&	--2.591 	&--3.221&	--2.811 	&	--2.237 	\\
		$	^	{	130	}	$Eu	&	1.031 \cite{G. Audi Chin. Phys. C 2017}	&$(1^+)\rightarrow(1/2^+,3/2^+)$&	2	&0.668&	--3.000 	&	--3.824  	&&	--3.267 	&	--2.695 	\\
		$	^	{	131	}	$Eu	&	0.963 	&$3/2^+\rightarrow0^+$&	2	&0.651&	--1.699 	&	--2.915 	&--2.578&	--2.458 	&	--1.840 	\\
		$	^	{	135	}	$Tb	&	1.203 	&$(7/2^-)\rightarrow0^+$&	3	&0.515&	--2.996 	&	--4.270  	&--4.222&	--3.806 	&	--3.212 	\\
		$	^	{	140	}	$Ho	&	1.104 	&$6^-,0^+,8^+\rightarrow(7/2^+)$&	3	&0.746&	--2.222 	&	--2.714	&--2.618&	--2.317 	&	--1.600 	\\
		$	^	{	141	}	$Ho	&	1.194 	&$(7/2^-)\rightarrow(0^+)$&	3	&0.803&	--2.387 	&	--3.772 	&--3.437&	--3.257 	&	--2.583 	\\
		$	^	{	141	}	$Ho$^m$	&	1.264 	&$(1/2^+)\rightarrow(0^+)$&	0	&0.715&	--5.137 	&	--6.052 	&--5.603&	--5.331 	&	--5.524 	\\
		$	^	{	144	}	$Tm	&	1.724 	&$(10^+)\rightarrow9/2^-\#$&	5	&0.031&	--5.569 	&	--4.117	&--5.288&	--4.687 	&	--4.991 	\\
		$	^	{	145	}	$Tm	&	1.754 	&$(11/2^-)\rightarrow0^+$&	5	&0.616&	--5.499 	&	--5.601	&--5.567&	--4.871 	&	--5.182 	\\
		$	^	{	146	}	$Tm	&	0.904 	&$(1^+)\rightarrow(1/2^+)$&	0	&0.748&	--0.810 	&	--0.814 	&--0.615&	--0.610 	&	--0.968 	\\
		$	^	{	146	}	$Tm$^m$	&	1.214 	&$(5^-)\rightarrow(1/2^+)$&	5	&0.748&	--1.137 	&	--1.190	&--1.083&	--0.896 	&	--0.737 	\\
		$	^	{	147	}	$Tm	&	1.072 	&$11/2^-\rightarrow0^+$&	5	&0.723&	0.587 	&	0.566 	&0.882&	0.614 	&	0.961 	\\
		$	^	{	147	}	$Tm$^m$	&	1.133 	&$3/2^+\rightarrow0^+$&	2	&0.709&	--3.444 	&	--3.212	&--2.588&	--2.859 	&	--2.183 	\\
		$	^	{	150	}	$Lu	&	1.285 	&$(5^-)\rightarrow(1/2^+)$&	5	&0.515&	--1.347 	&	--1.204 	&--1.173&	--1.132 	&	--0.965 	\\
		$	^	{	150	}	$Lu$^m$	&	1.305 	&$(1^+,2^+)\rightarrow(1/2^+)$&	2	&0.736&	--4.398 	&	--4.511	&--4.282&	--4.050 	&	--3.381 	\\
		$	^	{	151	}	$Lu	&	1.255 	&$11/2^-\rightarrow0^+$&	5	&0.521&	--0.896 	&	--0.903 	&--1.019&	--0.862 	&	--0.653 	\\
		$	^	{	151	}	$Lu$^m$	&	1.315 	&$3/2^+\rightarrow0^+$&	2	&0.747&	--4.796 	&	--4.626	&--4.355&	--4.150 	&	--3.471 	\\
		$	^	{	155	}	$Ta	&	1.466 	&$11/2^-\rightarrow0^+$&	5	&0.381&	--2.495 	&	--2.261	&--2.239&	--2.269 	&	--2.161 	\\
		$	^	{	156	}	$Ta	&	1.036 	&$(2^-)\rightarrow7/2^-\#$&	2	&0.680&	--0.826 	&	--0.623	&--0.153&	--0.624 	&	0.102 	\\
		$	^	{	156	}	$Ta$^m$	&	1.126 	&$(9^+)\rightarrow7/2^-\#$&	5	&0.405&	0.933 	&	1.323 	&1.487&	0.947 	&	1.371 	\\
		$	^	{	157	}	$Ta	&	0.946 	&$1/2^+\rightarrow0^+$&	0	&0.906&	--0.527 	&	--0.156	&--0.224&	--0.038 	&	--0.363 	\\
		$	^	{	159	}	$Re	&	1.816\cite{Blank B Prog. Part. Nucl. Phys. 2008} 	&$11/2^-\rightarrow0^+$&	5	&0.232&	--4.678 	&	--4.276	&&	--4.270 	&	--4.285 	\\
		$	^	{	159	}	$Re$^m$	&	1.816 	&$11/2^-\rightarrow0^+$&	5	&0.211&	--4.665 	&	--4.234 	&--4.372&	--4.269 	&	--4.283 	\\
		$	^	{	160	}	$Re	&	1.276 	&3/2$^+\rightarrow$0$^+$&	2	&0.670&	--3.163 	&	--3.051	&--3.441&	--2.841 	&	--2.101 	\\
		$	^	{	161	}	$Re	&	1.216 	&$1/2^+\rightarrow0^+$&	0	&0.908&	--3.357 	&	--3.295	&--2.830&	--2.895 	&	--3.018 	\\
		$	^	{	161	}	$Re$^m$	&	1.338 \cite{Blank B Prog. Part. Nucl. Phys. 2008}	&$11/2^-\rightarrow0^+$&	5	&0.227&	--0.678 	&	--0.390 	&--0.356&	--0.806 	&	--0.501 	\\
		$	^	{	164	}	$Ir	&	1.844\cite{Blank B Prog. Part. Nucl. Phys. 2008} 	&$(9^+)\rightarrow7/2^-$&	5	&0.169&	--3.947 	&	--3.937 	&--4.218&	--4.114 	&	--4.039 	\\
		$	^	{	165	}	$Ir$^m$	&	1.727 	&$(11/2^-)\rightarrow0^+$&	5	&0.170&	--3.433 	&--3.150 	&--3.172&	--3.408 	&	--3.266 	\\
		$	^	{	166	}	$Ir	&	1.167 	&$(2)^-\rightarrow(7/2^-)$&	2	&0.363&	--0.824 	&	--0.931	&--0.882&	--1.188 	&	--0.426 	\\
		$	^	{	166	}	$Ir$^m$	&	1.347 	&$(9)^+\rightarrow7/2^-$&	5	&0.190&	--0.076 	&	0.093 	&0.328&	--0.475 	&	--0.100 	\\
		$	^	{	167	}	$Ir	&	1.087 	&$1/2^+\rightarrow0^+$&	0	&0.894&	--1.120 	&	--1.016	&--0.627&	--0.865 	&	--1.076 	\\
		$	^	{	167	}	$Ir$^m$	&	1.262 	&$11/2^-\rightarrow0^+$&	5	&0.167&	0.842 	&	1.067	&0.621&	0.348 	&	0.798 	\\
		$	^	{	170	}	$Au	&	1.487 	&$(2)^-\rightarrow(7/2^-)$&	2	&0.241&	--3.487 	&	--3.658 	&--3.617&	--3.845 	&	--3.023 	\\
		$	^	{	170	}	$Au$^m$	&	1.767 	&$(9)^+\rightarrow7/2^-$&	5	&0.241&	--2.971 	&	--3.179 	&--2.803&	--3.333 	&	--3.118 	\\
		$	^	{	171	}	$Au	&	1.464 	&$1/2^+\rightarrow0^+$&	0	&0.872&	--4.652 	&	--4.755 	&--4.331&	--4.298 	&	-4.228 	\\
		$	^	{	171	}	$Au$^m$	&	1.718 	&$11/2^-\rightarrow0^+$&	5	&0.065&	--2.587 	&	--2.260 	&--2.936&	--3.026 	&	--2.777 	\\
		$	^	{	176	}	$Tl	&	1.278 	&$(3^-,4^-)\rightarrow(7/2^-)$&	0	&0.189&	--2.208 	&	--1.588 &--1.569&	--2.059 	&	--2.113 	\\
		$	^	{	177	}	$Tl	&	1.173 	&$(1/2^+)\rightarrow0^+$&	0	&0.498&	--1.174 	&	--0.699 	&--0.385&	--0.875 	&	--1.015 	\\
		$	^	{	177	}	$Tl$^m$	&	1.967\cite{Blank B Prog. Part. Nucl. Phys. 2008} 	&$(11/2^-)\rightarrow0^+$&	5	&0.022&	--3.346 	&	--3.086	&--3.453&	--4.227 	&	--3.972 	\\
		$	^	{	185	}	$Bi$^m$	&	1.625 	&$1/2^+\rightarrow0^+$&	0	&0.032&	--4.191 	&	--3.734	&--5.024&	--4.759 	&	--4.511 	\\
		\hline
	\end{tabular*}
\end{table*}

\begin{table*}[htb]
	\caption{The calculated results of the proton emission half-lives for deformed nuclei when the $l$ of $^{117}$La, $^{121}$Pr, $^{135}$Tb and $^{141}$Ho are taken as 3, 3, 4 and 4.}
	\label{tab2}
	\begin{tabular*}{18cm} {@{\extracolsep{\fill}} lcccccc}
		\hline
		{Nucleus}&$l$&&$\rm{log}_{10}{\emph{T}}_{1/2}$(s)&
		\\
		\cline{3-7}
		$^{A}Z$&&Exp&D-GLM&CPPM \cite{J. G. Deng Eur. Phys. J. A 2019}&UFM \cite{J. M. Dong Chin. Phys. C 2010}&UDLP \cite{Qi C Phys. Rev. C 2012}\\
		\hline	
		$	^	{	117	}	$La	&	3 	&	--1.664	&	--1.591  	&--1.587&--1.785&	--1.667  	\\
		$	^	{	121	}	$Pr	&	3 	&	--1.921	&	--1.742 	&--2.007&--2.158&--2.036 		\\
		$	^	{	135	}	$Tb	&	4 	&	--2.996	&	--3.252	&--2.948&--2.844&	--2.837  	\\
		$	^	{	141	}	$Ho	&	4 	&	--2.387	&	--2.760 	&--2.322&--2.214&	--2.310  	\\
		\hline
	\end{tabular*}
\end{table*}

Furthermore, for a more intuitive comparison of the experimental half-lives with the calculated ones, the deviations between the experimental proton emission half-lives in logarithmic form and the calculated ones obtained by using the above four models and/or formulae are plotted in Fig. \ref{fig3}. It is clearly seen that the majority of deviations are within $\pm 1$ on the whole, which indicates that the present estimates for the proton emission half-lives using the D-GLM are adequately credible. However, for some proton emitters, the calculated half-lives with the D-GLM differ the experimental ones by about one order of magnitude, corresponding to the well deformed cases of $^{117}$La, $^{121}$Pr, $^{135}$Tb, and $^{141}$Ho. Seeking the reasons why such a large deviation occurs is a vital topic. It is well known that the calculated results are sensitive to the released energy $Q_p$ and the angular momentum $l$. Consequently, the first possible reason is that the deviation between the results of experimental data and the calculated ones for the above four cases may be because of the uncertainties in measurements of the $Q_p$ values as well as it requires further investigation theoretically and measurements with high accuracy. Besides, for $^{117}$La, $^{121}$Pr, $^{135}$Tb and $^{141}$Ho, the proton emission half-lives in logarithmic form calculated by the D-GLM versus the $Q_p$ in the different cases of $l$ are presented in Fig. \ref{fig4}. From this figure, we can found that the half-lives increase by one order of magnitude or more for each increase in angular momentum when the values of $Q_p$ are same and $l>2$. Taking $^{117}$La as an example, the half-lives can reproduce experimental data well for $l=3$  rather than for $l=2$. In the same way, for the nuclei $^{121}$Pr, $^{135}$Tb and $^{141}$Ho, the experimental data are perfectly fitted, while the values of $l$ are taken as 3, 4 and 4. The values of $l$ have a higher consistency with the analytical results of Cheng \emph{et al.} \cite{J. H. Cheng Nucl. Phys. A 2020}, which means that the orbital angular momentum of $^{117}$La, $^{121}$Pr, $^{135}$Tb and $^{141}$Ho may be 3, 3, 4 and 4. Therefore, the second reason for large deviations is perhaps that the values of $l$ are uncertain.

In order to further verify the rationality of this conclusion, using the D-GLM, UDLP, the Coulomb and proximity potential model (CPPM) \cite{J. G. Deng Eur. Phys. J. A 2019}, and the unified fission model (UFM) \cite{J. M. Dong Chin. Phys. C 2010}, the proton emission half-lives are calculated while the $l$ of $^{117}$La, $^{121}$Pr, $^{135}$Tb and $^{141}$Ho are changed to 3, 3, 4 and 4. The detailed results are presented in Table. \ref{tab2}. From this table, we can see that the calculated results accurately match the experimental data. Moreover, the logarithmic form of the experimental proton emission half-lives and the calculated ones using the four methods mentioned above, are plotted in Fig. \ref{fig5}. The calculated results for $^{117}$La, $^{121}$Pr, $^{135}$Tb and $^{141}$Ho  with the angular momentum of $l=$2, 2, 3, 3 and $l=$3, 3, 4, 4, are presented in Fig. \ref{fig5}a and Fig. \ref{fig5}b, respectively. We note that the calculated results obtained by the D-GLM, CPPM, UFM, and UDLP more accurately reflect the experimental data after the values of $l$ are changed. The values of $l$ may provide a reference for future research. 

To globally comprehend the agreement between experimental data and calculated ones, we calculate the standard deviations obtained by the D-GLM, UDLP, and NG-N, denoted as $\sigma_{{\rm D-GLM}}$, $\sigma_{\rm UDLP}$ and  $\sigma_{\rm NG-N}$. All the detailed results of $\sigma$ are listed in Table.\ref{tab3}, which show that $\sigma_{{\rm D-GLM}}=0.533$, $\sigma_{{\rm GLM-Xiao}}=0.601$, $\sigma_{\rm UDLP}=0.471$ and $\sigma_{\rm NG-N}=0.515$. The results demonstrate that the proton emission half-lives calculated by the D-GLM have high accuracy.

\begin{figurehere}\centering
	\includegraphics[width=8cm]{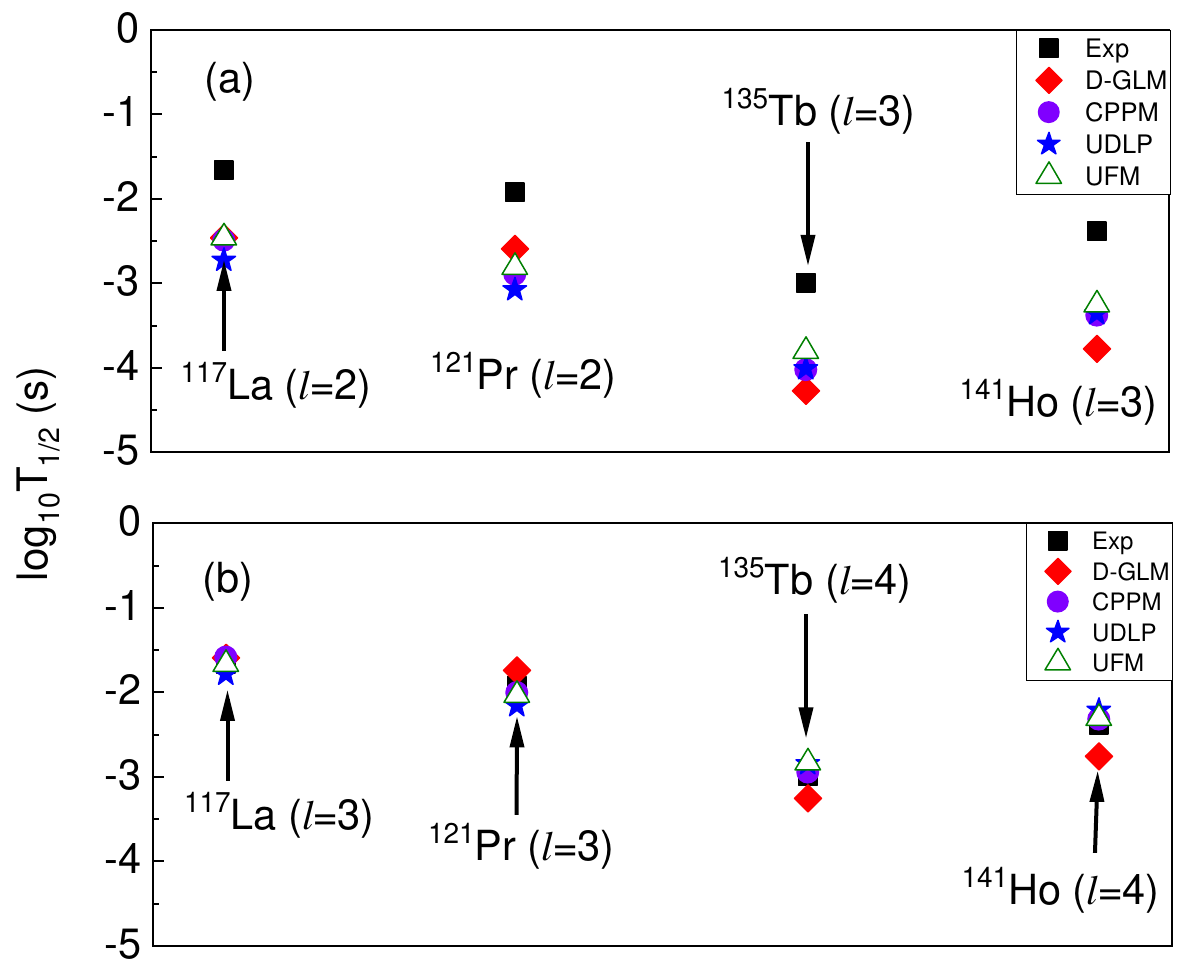}
	\figcaption{(color online) The comparison of the D-GLM and UDLP results with the experimental data for proton emitter of $^{117}$La, $^{121}$Pr, $^{135}$Tb and $^{141}$Ho for $l=$2, 2, 3, 3 $(a)$ and for $l=$3, 3, 4, 4 $(b)$  respectively.}
	\label{fig5}
\end{figurehere}

\begin{tablehere}
	\renewcommand\arraystretch{1.5}
	\tabcaption{The standard deviations $\sigma$ between experimental half-lives and calculated ones using the D-GLM, UDLP, and NG-N, respectively.} 
	\label{tab3} 
	\centering
	\begin{tabular*}{8.5cm} {@{\extracolsep{\fill}} ccccc}
		\hline
		Type & D-GLM&GLM-Xiao& UDLP & NG-N 
		\\  
		\hline 
		$\sigma$  & 0.533 &0.601&0.471
		& 0.515 \\	 
		\hline
	\end{tabular*}  
\end{tablehere}

\section{Summary}
\label{Sec.IV}
In conclusion, we offer a systematic study of the proton emission half-lives for 45 deformed proton emitters based on the D-GLM, which contains the deformed effect of Coulomb potential for the daughter nucleus. The present model contains only one adjustable parameter, i.e., the effective nuclear radius constant, determined to be $r_0=1.20$ fm, which is obtained by fitting 45 experimental data of the proton emission half-lives. We found that the calculated results could reproduce the experimental data well. Moreover, the relevance between the half-lives and $l$ for $^{117}$La, $^{121}$Pr, $^{135}$Tb and $^{141}$Ho was investigated, as well as corresponding possible reference values were proposed: $l=$3, 3, 4, 4. Future theoretical and experimental research could benefit from the useful knowledge that this work may offer.
\end{multicols}
\vspace{-1mm}
\centerline{\rule{80mm}{0.1pt}}
\vspace{2mm}

\begin{multicols}{2}

\end{multicols}

\end{CJK*}
\end{document}